\documentstyle[sprocl,epsfig,wrapfig]{article}

\def\Journal#1#2#3#4{{#1} {\bf #2}, #3 (#4)}

\def\NPPS{{\em Nucl. Phys.} (Proc. Suppl.)}
\def\PLB{{\em Phys. Lett.}  B}

\def\EPJC{{\em Eur. Phys. J.} C}
\def\JHEP{{\em JHEP}}

\def\be{\begin{equation}}
\def\ee{\end{equation}}
\def\bea{\begin{eqnarray}}
\def\eea{\end{eqnarray}}

\begin{document}
\vspace*{-3cm}
\vbox to 21mm {
\hbox to \textwidth{ \hsize=\textwidth
\hspace*{0pt\hfill} 
\vbox{ \hsize=58mm
{
\hbox{ MPI-PhE/2000-14 \hss}
\hbox{ June 15, 2000\hss } 
}
}
}
}
\title{EVENT SHAPES AND POWER CORRECTIONS IN e$^+$e$^-$ ANNIHILATION}

\author{O. BIEBEL}

\address{Max-Planck-Institut f\"ur Physik, F\"ohringer Ring 6,\\
80805 M\"unchen, Germany}

\maketitle\abstracts{ 
The effects of the hadronisation of partons on the distribution of event 
shape observables are associated with corrections which are suppressed 
by reciprocal powers of the energy scale of the process. The correction 
is determined by one non-calculable parameter $\alpha_0$ for which 
an universal value of $0.5 \pm 20\%$ is found from the investigation of 
the distribution of event shape observables and their mean values measured 
in e$^+$e$^-$ annihilation.
}

\section{Motivation}
When $\alpha_S$ is determined from event shapes in e$^+$e$^-$ 
annihilation the effects due to hadronisation need to be corrected. 
It yields a contribution to the overall error of $\alpha_S$ which 
is typically as large as the experimental systematics and the 
uncertainties associated with the choice of the scale. The error 
contribution might be alleviated by 
employing power corrections instead of phenomenological 
hadronisation models which need adjusting many parameters.

\section{Power Corrections to Mean Values}
Hadronisation is expected to cause corrections 
to measured observables which are suppressed by reciprocal powers 
of the energy scale of the process. In~\cite{bib-dokshitzer-webber} 
power corrections to the mean values of event shape observables 
are additive terms
\begin{equation}
\label{eqn-powcor}
\langle {\cal F}\rangle = \langle {\cal F}_{\mathrm{pert}}\rangle + 
                          \langle {\cal F}_{\mathrm{pow}}\rangle. 
\end{equation}
The correction $\langle {\cal F}_{\mathrm{pow}}\rangle \propto a_{\cal F}\cdot \alpha_0/\sqrt{s}$ 
depends on a calculable observable-specific parameter $a_{\cal F}$ and a single 
non-perturbative parameter $\alpha_0$ to be measured experimentally 
which is the mean of the strong coupling $\alpha_S$ between 0 and 2 GeV. 
This type of correction has been thoroughly investigated for the thrust 
($T$), the heavy jet mass ($M_H$), the total ($B_T$) and wide jet broadening 
($B_W$), and the C-parameter ($C$) observables for $\sqrt{s}=$14-202 GeV. 
The results,\cite{bib-l3note-2504,bib-delphi-conf343,bib-biebel-etal} updated 
for the corrected Milan factor,\cite{bib-dokshitzer} yield on average 
$\alpha_0(2 {\mathrm{~GeV}})= 0.49\pm 0.03$ (r.m.s.\ 0.07) supporting the 
universality of $\alpha_0$. The r.m.s., which is larger than 
the combined statistical, systematic and scale uncertainties, is partly due 
to the neglect of the $20\%$ uncertainty of the Milan factor.
The average of $\alpha_S(M_{\mathrm{Z}})$ is $0.116 \pm 0.004$ which agrees
with the world average.\cite{bib-bethke}
 
\subsection{Power Corrections to $y_3$}
\begin{figure}
\centerline{\psfig{figure=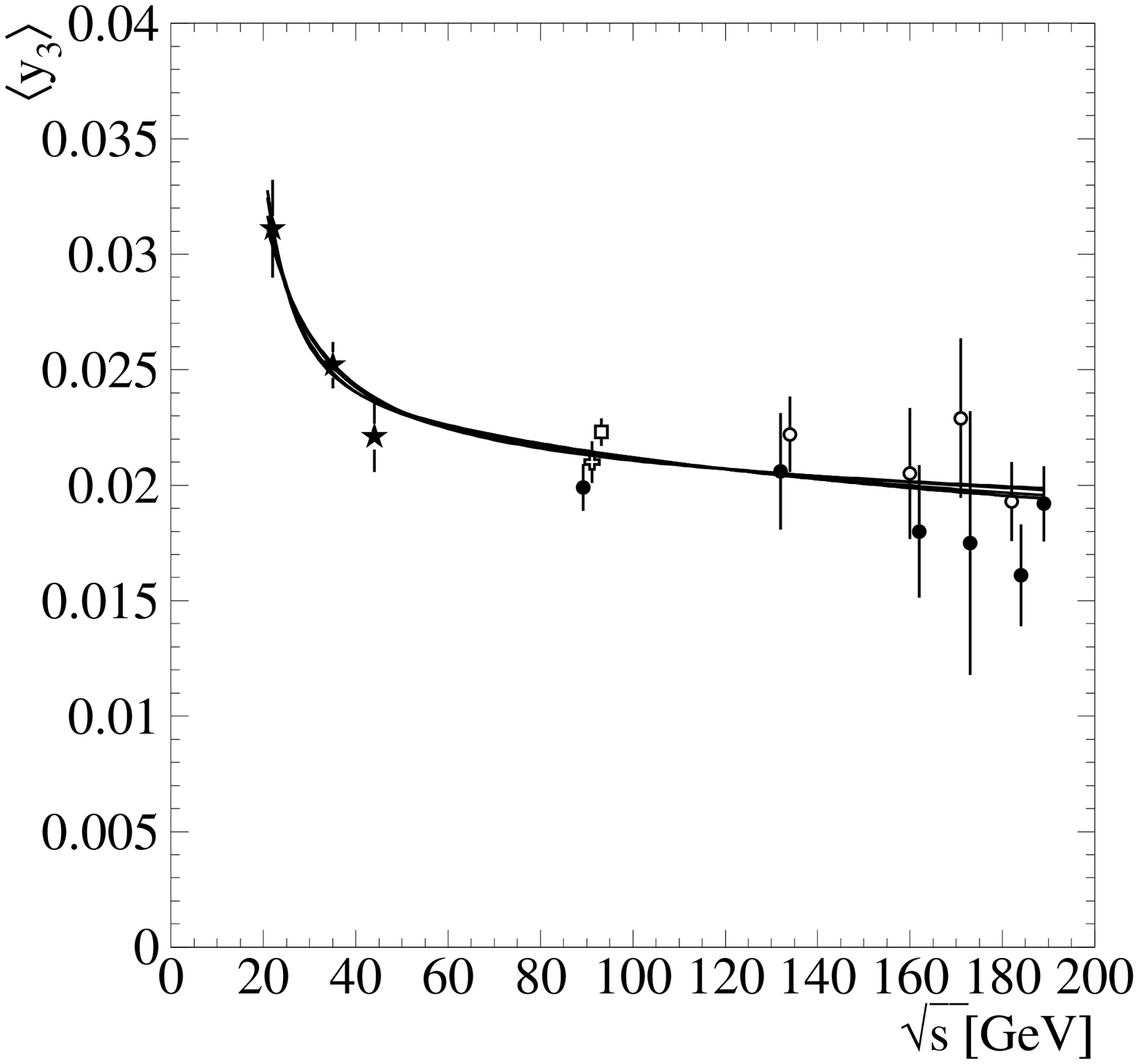,width=63mm}\hfill
            \psfig{figure=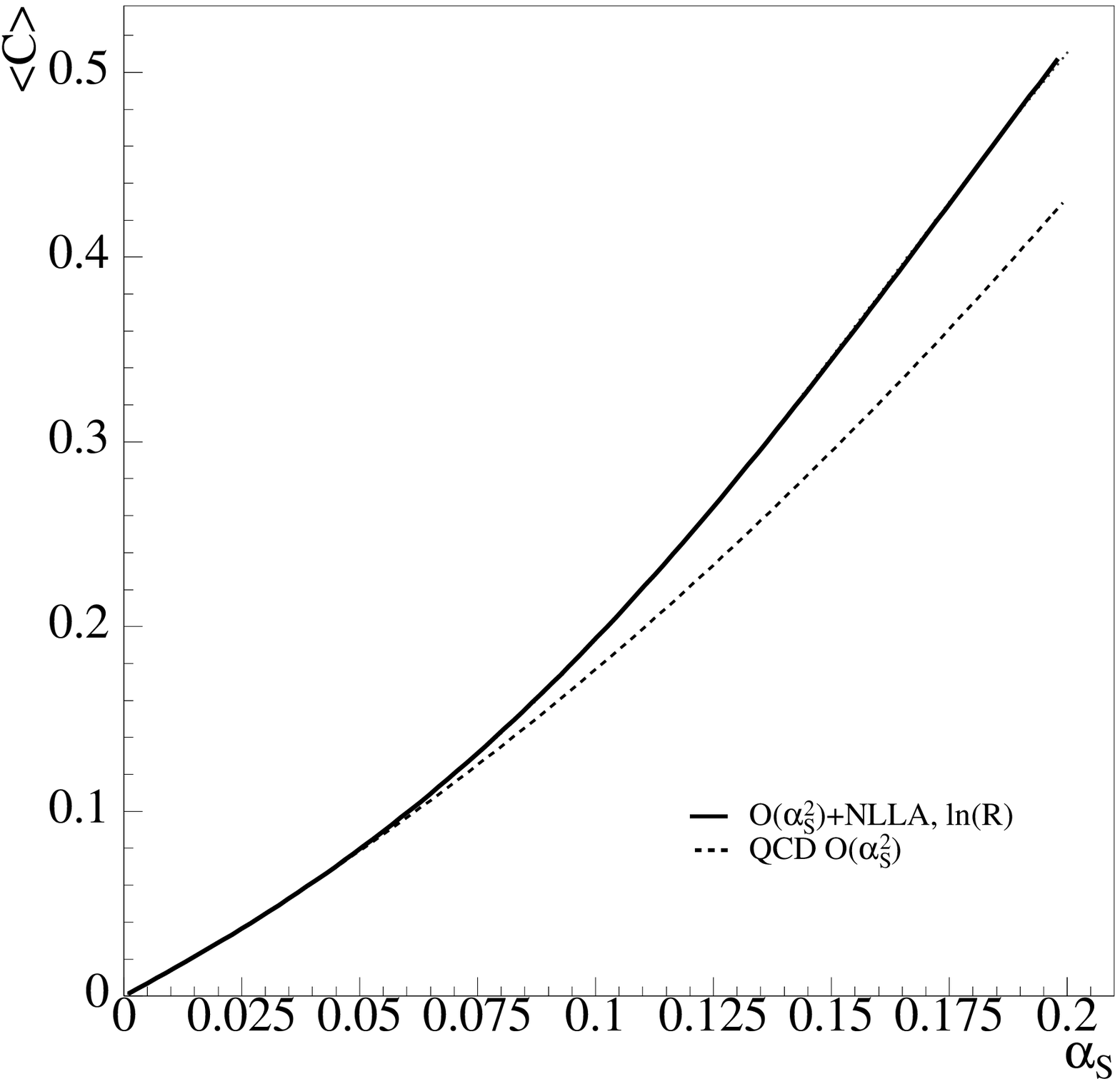,width=61mm}}
\caption{\label{fig-y3-powcor}\label{fig-resummed-means}
         Left:  Fits of $1/\sqrt{s}$, $\ln s/\sqrt{s}$, $1/s$, and 
                $\ln s/s$ power corrections to $\langle y_3 \rangle$.
         Right: Mean values of the C-parameter obtained using second 
                order ${\cal O}(\alpha_S^2)$ (dashed) and $\ln R$-matched 
                resummed NLLA plus ${\cal O}(\alpha_S^2)$ (solid line) 
                calculations.
}
\end{figure}

Many observables are subject to power corrections of the type $1/\sqrt{s}$ or 
$1/\sqrt{\ln s}$. One observable which is known to have a leading $1/s$ or $\ln s/s$ 
correction is the 2-3-jet flip, $y_3$, for the $k_{\perp}$ jet finder but the 
coefficient $a_y$ which determines the size of this correction is not known. 
Figure \ref{fig-y3-powcor} shows the fit results of power corrections of the 
type $1/\sqrt{s}$, $\ln s/\sqrt{s}$, $1/s$, and $\ln s/s$ to the mean of $y_3$.
All fits give $\chi^2/$d.o.f.\ of about 1 but the high values 
$\alpha_S(M_{\mathrm{Z}})=0.144\pm 0.011_{\mathrm{expt'l}}$ disfavour 
the $1/\sqrt{s}$-type corrections. The $1/s$ correction yields 
$a_y = -0.49 \pm 0.37_{\mathrm{expt'l}}$ compatible with zero 
for $\alpha_S(M_{\mathrm{Z}})=0.124\pm 0.004_{\mathrm{expt'l}}$ and 
for the first moment of $\alpha_S$ in the range of 0 through 2 GeV,
$\alpha_1=0.26 \pm 0.02_{\mathrm{expt'l}}$.

\subsection{Resummed Predictions of Mean Values}
Usually all investigations used the second order calculations for the
mean values of the event shape observables but the matched resummed
and fixed order calculations for the distributions (see sect.~\ref{sec-distributions}). 
Resummed predictions for the mean values should give a better description of the 
dominating contribution from the 2-jet region to the mean value. The result
of matching the resummed NLLA prediction for the mean value with the fixed
order calculation is exemplified in figure~\ref{fig-resummed-means} for
the C-parameter. 
A comparable change of the ${\cal O}(\alpha_S^2)$ result is found for 
$\langle M_H\rangle$. The difference is about twice as large for the 
$\langle B_W\rangle$ while it is negligible for $\langle 1-T\rangle$ 
and $\langle B_T\rangle$. Fitting the data, $\alpha_0$ turns out to 
be 5-10\% lower ($-40\%$ for $B_W$) and $\alpha_S(M_{\mathrm{Z}})$ to 
be 1-10\% lower if using the $\ln R$-matched resummed and fixed order 
calculations.

\section{Power Corrections to Higher Moments}
\begin{figure}
\centerline{\psfig{figure=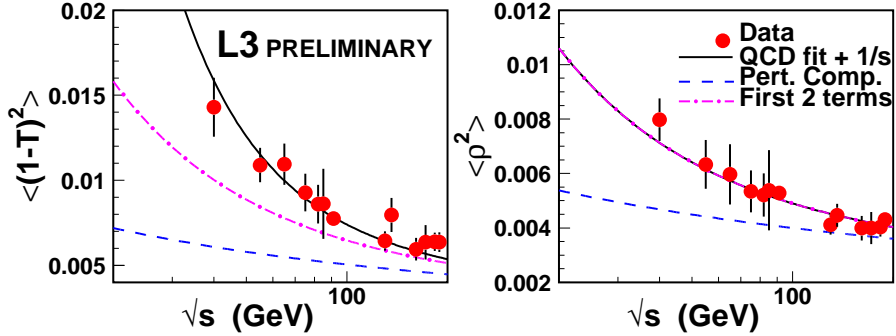,width=118mm}}
\caption{\label{fig-2ndmoments}
         Power corrections to second moment of thrust (left) and heavy
         jet mass (right). 
}
\end{figure}
An straight forward extension of Eq.~(\ref{eqn-powcor}) to the second moment
of the event shapes yields
$
\langle {\cal F}^2\rangle = \langle {\cal F}^2_{\mathrm{pert}}\rangle + 
                          2 \langle {\cal F}_{\mathrm{pert}}\rangle\cdot
                            \langle {\cal F}_{\mathrm{pow}}\rangle +
                           {\cal O}(1/s).
$
For the second moment of the thrust observable, however, a 
$1/(\sqrt{s})^3$ power correction is expected in the 2-jet 
region.\cite{bib-gardi} 
The investigations~\cite{bib-l3note-2504} exemplified in 
figure~\ref{fig-2ndmoments} show that $\langle 1-T\rangle$, 
$\langle B_T\rangle$, and $\langle C\rangle$ require a large $1/s$ 
term which is not necessary for $\langle \rho^2\rangle\equiv 
\langle M_H^2/s\rangle$ and $\langle B_W\rangle$. 

To suppress the $\langle {\cal F}_{\mathrm{pow}}\rangle$ term in the 
formula of the second moment
the study of the variance has been proposed.\cite{bib-webber} 
With no other data available the variance of the C-parameter, 
$\sigma^2_C \equiv \langle C^2\rangle - \langle C\rangle^2 =
0.034 \pm 0.010$ has been calculated from the distributions 
measured at 91 GeV using error propagation to assess the total 
error. Using the second order prediction, 
$\sigma_C^2 \approx 0.387\alpha_S + 0.0435\alpha_S^2$,
and neglecting the $1/s$ correction the variance yields a very
low value $\alpha_S(M_{\mathrm{Z}})=0.09\pm 0.03$ which could be 
due to the C-parameter spectrum which does not vanish at the 
boundary of the 3-jet phase space.\cite{bib-korchemsky} In general, 
more complete predictions are required to make use of the higher 
order moments.

\section{Power Corrections to Differential Distributions}
\label{sec-distributions}
Power corrections can also be applied to the differential distributions 
of event shapes by shifting the resummed plus ${\cal O}(\alpha_S^2)$
prediction. This has been investi- 
%
\begin{wrapfigure}{r}{60mm}
\centerline{\psfig{figure=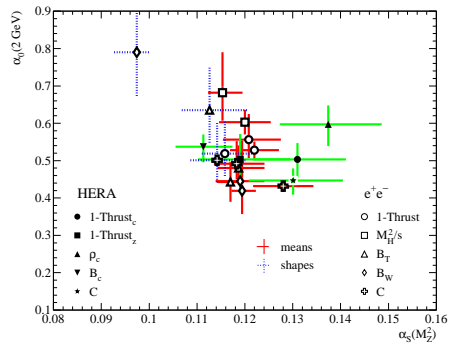,width=59mm}}
\caption{\label{fig-as-vs-a0}
         Fit results of $\alpha_0$ and $\alpha_S(M_{\mathrm{Z}})$.
}
\end{wrapfigure}
%
gated 
by several groups using $T$ and $M_H$. With the reanalysed JADE data at 
35 and 44 GeV also power corrections to the
$C$ and jet broadening distributions became possible which showed the 
necessity of squeezing the predicted distributions for the latter in 
addition to the shift.\cite{bib-biebel-etal} The fit results from $T$,
$B_T$, $B_W$, and $C$, updated for the corrected Milan factor, yield 
on average $\alpha_0 = 0.57 \pm 0.09$ (r.m.s.\ 0.12) and 
$\alpha_S(M_{\mathrm{Z}})=0.107 \pm 0.006$. The $\chi^2/$d.o.f.\ of the 
fits is about unity but for $B_W$ which yields $\alpha_0=0.79$ too high
and $\alpha_S(M_{\mathrm{Z}})=0.097$ too low. 

These fits of the event shape distributions exclude the extreme 2-jet 
region. Extending the power corrections with a shape function~\cite{bib-korchemsky2} 
a fit over the whole distribution is possible.

\section{Conclusions}
Figure~\ref{fig-as-vs-a0} summarises the results of the fits of 
$\alpha_0(2 {\mathrm{~GeV}})$ and $\alpha_S(M_{\mathrm{Z}})$ from 
the mean values and from the distributions of event shapes in e$^+$e$^-$ 
annihilation. These results agree with those from studies of 
event shapes in ep scattering.\cite{bib-H1} In all 
these investigations power corrections prove to be a useful 
description of the hadronisation effects and the single 
non-perturbative parameter $\alpha_0(2 {\mathrm{~GeV}})$ 
assumes an universal value of about $0.5 \pm 20\%$.

\end{document}